\documentclass[a4paper,3p]{elsarticle}
\pdfoutput=1

\usepackage{lineno}
\modulolinenumbers[5]
\usepackage[colorlinks=true,linkcolor=blue,citecolor=blue,urlcolor=blue]{hyperref}
\journal{}

\biboptions{sort&compress}

\usepackage[cmex10]{amsmath}
\usepackage{amssymb}
\usepackage{mdwlist}
\usepackage{upgreek}
\usepackage[section]{placeins}

\usepackage{pinlabel}
\newcommand*{\labfont}{\fontfamily{phv}\selectfont}

\usepackage{booktabs}
\usepackage{tabulary}

\usepackage{listings}             % Include the listings-package
\usepackage[numbered]{matlab-prettifier}
\lstMakeShortInline[style=Matlab-editor]"

\hypersetup{pdfauthor={JM O'Toole and GB Boylan},
  pdftitle={NEURAL: quantitative features for newborn EEG using Matlab},
  pdfkeywords={neonate, preterm infant, electroencephalogram, quantitative analysis, feature
    extraction, spectral analysis, inter-burst interval}}

\newcommand{\ee}{\text{e}}
\newcommand{\jj}{\text{j}}

\newcommand{\ver}{v0.3.1}

\begin{document}
\lstset{language=Matlab}

\begin{frontmatter}
  \title{NEURAL: quantitative features for newborn EEG using Matlab}
  
  \author[INFANT]{John~M.~O'~Toole\corref{mycorrespondingauthor}}
  \cortext[mycorrespondingauthor]{Corresponding author}
  \ead{jotoole@ucc.ie}

  \author[INFANT]{Geraldine~B.~Boylan}
  \ead{G.Boylan@ucc.ie}

  \address[INFANT]{Neonatal Brain Research Group, Irish Centre for Fetal and
    Neonatal Translational Research (INFANT), University College Cork, Ireland}

  \begin{abstract}
    \textit{Background}: For newborn infants in critical care, continuous monitoring of
    brain function can help identify infants at-risk of brain injury.
    Quantitative features allow a consistent and reproducible approach to EEG analysis,
    but only when all implementation aspects are clearly defined.
    \textit{Methods}: We detail quantitative features frequently used in neonatal EEG
    analysis and present a Matlab software package together with exact implementation
    details for all features.
    The feature set includes stationary features that capture amplitude and frequency
    characteristics and features of inter-hemispheric connectivity.
    The software, a Neonatal Eeg featURe set in mAtLab (NEURAL), is open source and freely
    available.
    The software also includes a pre-processing stage with a basic artefact removal
    procedure.  
    \textit{Conclusions}: NEURAL provides a common platform for quantitative analysis of
    neonatal EEG.
    This will support reproducible research and enable comparisons across independent
    studies.
    These features present summary measures of the EEG that can also be used in automated
    methods to determine brain development and health of the newborn in critical care.
  \end{abstract}

\end{frontmatter}

\vspace{1em}
\emph{Abbreviations:} EEG, electroencephalogram; NEURAL, neonatal EEG feature set in
{\sc Matlab}; NICU, neonatal intensive care unit; aEEG, amplitude-integrated EEG;
rEEG, range EEG; PSD, power spectral density; BSI, brain symmetry index; FIR, finite
impulse response; IIR, infinite impulse response; DFT, discrete Fourier transform; CC,
correlation coefficient.  

%\linenumbers
\section{Introduction}
\label{sec:introduction}

Electroencephalography (EEG) is used in the neonatal intensive care environment to monitor
brain function of critically-ill newborns. 
This non-invasive and portable technology provides continuous assessment of cortical
function at the cot side, with little interruption to standard clinical care.  
Specialists are required to visually interpret the EEG to evaluate brain health, by
identifying seizures if present \cite{Boylan2013}, assessing brain maturation
\cite{Andre2010}, or grading brain injury \cite{Lloyd2016}.  
Yet in many neonatal intensive care units (NICUs), provision for continuous reporting on
the EEG of multiple infants is constrained by the availability of the specialist.    
Addressing this limitation, many NICUs use an amplitude-integrated EEG (aEEG) device
instead of the EEG. 
The aEEG presents a band-pass filtered and time-compressed version of 1 or 2 channels of
EEG \cite{Boylan2008a,Tao2010}. 
Because of the time-compression, a long duration EEG (approximately 6 hours) is summarised
in 1 page, making reviewing an easier task that is often preformed by non-EEG specialists,
such as the treating clinician. 
Yet the time-compression destroys much of the detail of the EEG waveform and many
important clinical features, such as short-duration seizures, are not presented on the
aEEG \cite{Boylan2008a}. In addition, artefacts can falsely enhance baseline activity or
may be misinterpreted as seizures \cite{Hagmann2006,Suk2009,Marics2013}.  

Quantitative EEG analysis provides an alternative to visual interpretation, with specific
advantages.  
First, quantitative analysis provides consistency without the varying degrees of
inter-rate agreement associated with visual interpretation \cite{Stevenson2015}.
Second, quantitative analysis can uncover attributes not accessible with visual analysis
alone, such as measures of connectivity \cite{Tokariev2012,Omidvarnia2013b}.
Third, quantitative analysis can facilitate reproducible research for clinical,
scientific, and engineering studies.
And last, quantitative analysis is a necessary component for developing fully automated
methods of EEG analysis \cite{Greene2008,Temko2010,OToole2016b}. 
This is of particular importance as automated analysis of the EEG addresses the need to
provide continuous, around-the-clock EEG reporting in the
NICU---something not possible with visual interpretation alone.  

Quantitative features describe many aspects of newborn EEG, including sleep cycles
\cite{scher1994comparisons,Duffy2003,Paul2003,Scher2004a}, normative ranges
\cite{West2006,Victor2005b,Pereda2006,Okumura2006,Korotchikova2009,Gonzalez2011,Thorngate2013},
association with clinical outcomes
\cite{Inder2003,Korotchikova2011,Williams2012,Saito2013,Schumacher2013a}, and functional
maturation
\cite{Burdjalov2003,Niemarkt2011,O'Reilly2012,Meijer2014,Shany2014,Saji2015,OToole2016b}.   
Not strictly defined, the term \emph{quantitative EEG features} typically refers to basic
signal processing measures of frequency and amplitude. 
Most measures assume the signal is stationary and therefore rely on short-time analysis to
circumvent this assumption.  
Although the features frequently appear in EEG literature, implementation details are
often omitted which makes comparison between independent studies difficult, as different
implementations will generate different estimates.  
The goal of this paper, therefore, is to present a clearly defined feature set with a
software implementation that is both open source and freely available.  
We hope the availability of this feature set will enhance comparisons between different
studies, support reproducible research, and enhance quantitative analysis of neonatal EEG.

%-------------------------------------------------------------------------------
% summary of the PACKAGE	
% -------------------------------------------------------------------------------

\section{NEURAL: software package for Matlab}
\label{sec:queen:-softw-pack}

The software package \emph{NEURAL}---a \underline{n}eonatal \underline{E}EG
feat\underline{ur}e set in m\underline{a}t\underline{l}ab (\ver)---runs within the {\sc
  Matlab} environment (The MathWorks, Inc., Natick, Massachusetts, United States).  
Details on how to install and setup the {\sc Matlab} code is described in the "README.md"
file accompanying the package \cite{OToole2016c}.
The package can generate multiple features on continuous, multi-channel EEG recordings. 
Features are defined specifically for neonatal EEG, including preterm infants, but could
also be applied to paediatric and adult EEG.  

Features are grouped into 4 categories:
\begin{itemize*}
\item \textbf{amplitude}: absolute amplitude and envelope of EEG signal, range EEG (similar to
  amplitude-integrated EEG); full list in Table~\ref{tab:feats_reeg}.  
\item \textbf{spectral}: spectral power (absolute and relative), spectral entropy (Wiener
  and Shannon), spectral differences, spectral edge frequency, and fractal dimension;
  full list in Table~\ref{tab:feats_spec}.  
\item \textbf{connectivity}: coherence, cross-correlation, and brain symmetry index; full
  list in Table~\ref{tab:feats_conn}.  
\item \textbf{inter-burst interval}: summary measures based on the inter-burst interval
  annotation (only relevant to preterm infants $<$32 weeks of gestational age); full
  list in Table~\ref{tab:feats_ibi}.  
\end{itemize*}

All features are listed in Tables~\ref{tab:feats_spec}--\ref{tab:feats_ibi} and in file
"all_features_list.m".  
Parameters, for both the features and pre-processing stage, are set in the
"neural_parameters.m" file. 
For example, the low-pass filter and new sampling frequency are set in
"neural_parameters.m" as
\begin{lstlisting}[style=Matlab-editor]
%% PREPROCESSING (lowpass filter and resample)
LP_fc=30;  % low-pass filter cut-off
Fs_new=64; % down-sample to Fs_new (Hz)
\end{lstlisting}
Default values set in this file are reported throughout. 

Unless otherwise stated, we use infinite impulse response (IIR) filters for all band-pass
filtering operations, for both the artefact removal and feature estimation procedures.
These filters are generated with a 5th order Butterworth design and implemented using the
forward--backwards approach to generate a zero-phase response \cite{oppen_dsp}.

\section{Pre-processing: Artefact Removal}
\label{sec:analys-cont-multi}

If required, the raw EEG can be pre-processed and then saved as Matlab files.
Pre-processing includes a simple artefact removal procedure followed by low-pass filtering
and downsampling. 
The function "resample_savemat.m" performs the pre-processing.
The artefact removal process removes major artefacts only, such as electrode coupling or
large-amplitude segments caused by movement; this process is detailed in the following and
includes similar procedures to those described in \cite{OToole2016b}.
If the EEG is still contaminated by other artefacts, such as respiration and heart rate,
other methods may also be needed \cite{Stevenson2014}.   
After artefact removal, all EEG channels are low-pass filtered (with a default frequency
of 30~Hz; parameter "LP_fc") using a finite-impulse response (FIR) filter. 
This filter is designed using the window method with a Hamming window of length 4,001
samples.
The EEG is then downsampled to a lower frequency (default 64~Hz; parameter "Fs_new").  

% \subsection{Removing Major Artefacts}
% \label{sec:pre-processing}
To enable the artefact removal process, set "REMOVE_ART=1" if
using function "resample_savemat.m"; otherwise, to turn off set "REMOVE_ART=0".
The process can also be implemented directly using the "remove_artefacts.m" function---see
Section~\ref{sec:removing-artefacts} for an example on how to use this function.  The
following details 5 stages of the artefact removal process.  The process requires both the
bipolar (set with "BI_MONT") and referential montage of EEG.  

The first stage is applied to the $(M+1)$-channel referential montage. 
Each channel $x_m[n]$, for $m=1,2,\ldots,M,M+1$, is length-$N$.  
\begin{enumerate*}
\item Improper electrode placement: 
  \begin{enumerate*}
  \item filter each channel $x_m[n]$ with 0.5--20~Hz bandpass filter;
  \item generate correlate coefficient $r_{pq}$ between channel $p$ and $q$, for
    $p,q=1,2,\ldots,M$, with $p\neq q$;
  \item average $r_{pq}$ over $q$, that is let $\bar{r}_p=1/(M-1) \sum_{q=1;\, q\neq p}^{M}
    r_{pq}$ for all $p$;
  \item for $\bar{r}_p<T_{\textrm{cc}}$ then remove channel $p$ from further analysis
    (default $T_{\textrm{cc}}=0.15$; parameter "ART_REF_LOW_CORR")
\end{enumerate*}

  The next stage uses the $M$ channel bipolar montage, with $x_m[n]$ now
  representing each bipolar channel.  

\item Electrode coupling:
  \begin{enumerate*}
  \item filter each channel $x_m[n]$ with 0.5--20~Hz bandpass filter;
  \item generate total power for the $m$th channel $P_m = 1/N \sum_{n=0}^{N-1} | x_m[n] |^2$;
  \item let $l$ denote the index of left-hemisphere channels only, where $\{l\}$ is a
    length-$M/2$ subset of $\{m\}$;  
  \item let $T_{\textrm{left}}=\textrm{median}(P_l)/4$
  \item if $P_l < T_{\textrm{left}}$, remove $l$th channel;
  \item repeat previous 3 steps for right-hemisphere channels.  
  \end{enumerate*}

  For the remaining stages process each bipolar channel $x[n]$ separately.
  
\item Continuous rows of zeros (on some EEG devices zeros replace EEG when testing for
  electrode impedance):
  \begin{enumerate*}
  \item identify segments $x[n]=0$ for $n=n+1,n+2,\ldots,n+L$; 
  \item if $L>T_{\textrm{min\_dur}}$, then remove (default $T_{\textrm{min\_dur}}=1$
    second; parameter "ART_ELEC_CHECK").
  \end{enumerate*}

\item High-amplitude artefacts:
  \begin{enumerate*}
  \item filter $x[n]$ with a [0.1--40]~Hz bandpass filter;
  \item generate the analytic associate of $x[n]$ as
    $z[n]=x[n]+\jj \mathcal{H} \{x[n]\}$, using the Hilbert transform $\mathcal{H}$.   
  \item if $|z[n]|>T_{\textrm{amp}}$, then remove this segment (default $T_{\textrm{amp}}=1500 \mu V$, suitable for preterm infants $<$32 weeks of gestation; parameter
    "ART_HIGH_VOLT")
  \item apply a collar around this segment and remove; default collar length is 10
    seconds (parameter "ART_TIME_COLLAR").  
  
\end{enumerate*}
\item Row of constant values or impulse-like activity (sudden jumps in
  amplitude):
  \begin{enumerate*}
  \item calculate the derivative of $x[n]$ using the forward-difference approximation
    $x'[n]=x[n+1]-x[n]$;
  \item identify $x'[n]=C$ for $n=n+1,n+2,\ldots,n+L$, where $C$ is a constant;
  \item if $L>T_{\textrm{min\_dur}}$, then remove (default $T_{\textrm{min\_dur}}=0.1$
    seconds; parameter "ART_DIFF_MIN_TIME");
  \item if $|x'[n]|>T_{\textrm{amp\_diff}}$, then remove (default
    $T_{\textrm{amp\_diff}}=200 \mu V$; parameter "ART_DIFF_VOLT").  
  
\item apply a collar around both artefacts if present and remove; default length of
    collar is 0.5 seconds (parameter "ART_DIFF_TIME_COLLAR")
  \end{enumerate*}
\end{enumerate*}

For artefacts identified on a single channel (preceding stages 3--5), the same segments
are removed over all channels.  
Identified artefacts are replaced by either zeros, linear interpolation, or cubic spline
interpolation; default is cubic spline interpolation, parameter
"FILTER_REPLACE_ARTEFACTS='cubic_interp'". 
For stages 1--2, if channels are identified as artefacts they are then removed from
further analysis.  

\section{Features}
\label{sec:features}

All features are estimated using the bipolar montage of the EEG. 
Many features are estimated within four different frequency bands of the EEG.  
Default values for these bands are [0.5--4; 4--7; 7--13; 13--30]~Hz, recommended for
infants $\geq 32$ weeks, or [0.5--3; 3--8; 8--15; 15--30]~Hz for preterm infants $<32$
weeks of gestation \cite{OToole2016b}. 
These bands are set with the "FREQ_BANDS" parameter and can also be set individually for
the specific feature, as we describe in the following.

\subsection{Amplitude}
\label{sec:amplitude}

Each EEG channel $x[n]$ is filtered into the four frequency bands to generate $x_i[n]$
($i=1,2,3,4$).
These bands can be set with the parameter "feat_params_st.amplitude.freq_bands".  
Amplitude is quantified by signal power and signal envelope.  
Signal power ($A_{\textrm{power}}^i$) is calculated as the mean value, over time ($n$), of
$|x_i[n]|^2$ and amplitude of signal envelope ($E_{\textrm{mean}}^i$) is calculated as the
mean value of envelope $e_i[n]$, with
\begin{equation}
  \label{eq:7}
  e_i[n]=\left| x_i[n] + \jj \mathcal{H}\{ x_i[n] \} \right|^2
\end{equation}
where $\mathcal{H}$ represents the discrete Hilbert transform implemented according to the
definition described by Marple \cite{marple}. 

Measures of variability of the EEG about a mean value are estimated by the standard
deviation ($A_{\textrm{sd}}^i$) of $x_i[n]$ and standard deviation of the envelope
$e_i[n]$. Skewness and kurtosis ($A_{\textrm{sk}}^i$ and $A_{\textrm{ks}}^i$) of $x_i[n]$
are calculated to estimate a non-Gaussian processes. 
Moments of the probability distribution (mean, standard deviation, skew, and kurtosis) are
defined in ~\ref{sec:moments-stat-rand}.  

Also included are features of the range-EEG (rEEG) \cite{O'Reilly2012}. 
rEEG was proposed as an alternative to amplitude-integrated EEG (aEEG) as there is no
clear definition of aEEG in the literature and most EEG machines implement different
versions of the aEEG algorithm \cite{Zhang2013}.
Although another measure of amplitude, the rEEG estimates a peak-to-peak measure of
voltage and therefore differs to the previous measures.  

We can calculate features of rEEG using either the full-band signal $x[n]$ or on the
individual frequency bands $x_i[n]$, which gives more flexibility than aEEG which uses a
fixed pass-band of 2--15~Hz \cite{Boylan2008a,Zhang2013}. 
(The passband 2--15~Hz is suboptimal for neonatal EEG analysis given the predominance of
delta power in neonatal EEG.)
Over a short-time windowed segment, the difference between the maximum and minimum is
generated
\begin{equation}
  \label{eq:16}
  r_i[l] = \max(x_i[n]w[n-lK]) - \min(x_i[n]w[n-lK])
\end{equation}
for window $w[n]$ (default rectangular window of length 2 seconds, parameters
"feat_params_st.rEEG.window_type" and "feat_params_st.rEEG.L_window") and time-shift
factor $K$ (parameter "feat_params_st.rEEG.overlap") related to the percentage overlap $H$
and window length $M$ as $K=\lceil M(1 - H/100) \rceil$. 
When plotting, the rEEG is transformed to a linear--log amplitude as follows
\begin{equation}
  \label{eq:17}
    r_i[l] =
  \begin{cases}
    \frac{50}{\log 50} \log r_i[l] & \textrm{if } r_i[l]>50 \\
    r_i[l] & \textrm{otherwise} \\
  \end{cases}
\end{equation}

Multiple features are used to summarise $r_i[l]$: mean $(R_{\textrm{mean}}^i)$ and median
$(R_{\textrm{median}}^i)$ as measures of central tendency; the 5th
$(R_{\textrm{lower}}^i)$ and 95th $(R_{\textrm{upper}}^i)$ percentiles to represent the
lower and upper margins; standard deviation $(R_{\textrm{sd}}^i)$, the coefficient of
variation $(R_{\textrm{cv}}^i=R_{\textrm{sd}}^i/R_{\textrm{mean}}^i)$, the difference
between the upper and lower margins
($R_{\textrm{bw}}^i=R_{\textrm{upper}}^i - R_{\textrm{lower}}^i)$ as measures of spread,
and a measure of symmetry defined as
\begin{equation}
  \label{eq:18}
  R_{\textrm{symm}}^i = \frac{(R_{\textrm{upper}}^i-R_{\textrm{median}}^i)-(R_{\textrm{median}}^i - R_{\textrm{lower}}^i)}{R_{\textrm{bw}}^i}
\end{equation}
where $R_{\textrm{symm}}^i$ ranges from $-1$ to $1$ with values close to $0$ indicating
symmetry and values close to $\pm1$ indicating asymmetry of the rEEG.

\subsection{Frequency}
\label{sec:spectral}

The following measures quantify spectral characteristics. 
First we present 3 ways to estimate the power spectral density (PSD) for EEG signal $x[n]$
of length-$N$ with sampling frequency $f_s$~Hz. 
These different PSD estimates are used in different features. 
The first PSD estimate is the periodogram $V[k]$ using a rectangular window
\begin{equation}
  \label{eq:2}
  V[k]=\frac{1}{Nf_s}\left| \sum_{n=0}^{N-1} x[n]\ee^{-\jj 2\uppi kn/N} \right|^2.  
\end{equation}
The second PSD estimate is the Welch periodogram $P[k]$
\begin{equation}
  \label{eq:8}
  P[k]=\frac{1}{LMUf_s}\sum_{l=0}^{L-1} \left| \sum_{n=0}^{M-1} x[n]w[n-lK]\ee^{-\jj 2\uppi kn/M} \right|^2
\end{equation}
where $w[n]$ is the analysis window of length $M$ with energy
$U=1/M \sum_{n=0}^{M-1} |w[n]|^2$; default settings apply a Hamming window of length 2
seconds, set with parameters "feat_params_st.spectral.window_type" and
"feat_params_st.spectral.L_window". 
The time-shift factor $K$ is related to the percentage overlap $H$ and window length $M$,
as $K=\lceil M(1 - H/100) \rceil$. 
Lastly, $L=\lfloor (N+K-M)/K \rfloor$ is the number of segments.  
Default value for $H$ is $50$\%, set with parameter "feat_params_st.spectral.overlap".  
And the third PSD estimate is a variant of the Welch periodogram, defined as 
\begin{equation}
  \label{eq:9}
  P_{\textrm{med}}[k]= \underset{l\in [0,L-1]}{\mathrm{median}}\left\{\frac{1}{MUf_s}
    \left| \sum_{n=0}^{M-1} x[n]w[n-lK]\ee^{-\jj 2\uppi kn/M} \right|^2 \right\}
\end{equation}
This estimate, which we refer to as the robust-PSD estimate, simply replaces the averaging
procedure of the Welch periodogram in \eqref{eq:8} with the median operator to generate a
more robust spectral estimate \cite{Zoubir2012}.

For features of absolute and relative spectral power at the $i$th frequency band, we use
the periodogram $V[k]$ in \eqref{eq:2}:
\begin{align}
  \label{eq:10}
  S_{\textrm{abspow}}^i & =\frac{s[k]f_s}{N}\sum_{k=a_i}^{b_i} V[k] \\
  \label{eq:1001}
  S_{\textrm{relpow}}^i & =\frac{\sum_{k=a_i}^{b_i} V[k]}{\sum_{k=a_1}^{b_4} V[k] } 
\end{align}
where $[a_i,b_i]$ represents the discrete frequency range of the $i$th frequency band.
The parameter $s[k]$ is a scaling factor, with $s[k]=2$ for
$k=1,2,\ldots,\lceil N/2 \rceil -1$ and $s[k]=1$ for the DC ($k=0$) and Nyquist frequency
($k=N/2$, for $N$ even only). 
This scaling factor is applied to conserve total power in the spectrum when using only the
positive frequencies of $V[k]$.

Spectral entropy measures are estimated using Wiener entropy, also known as
\emph{spectral flatness}, and Shannon entropy:
\begin{align}
  \label{eq:11}
  F_{\textrm{wiener}}^i & = \frac{\exp{\left( 1/L_i\sum_{k=a_i}^{b_i} \log P[k]\right)}}%
                          {1/L_i\sum_{k=a_i}^{b_i} P[k]} \\
  F_{\textrm{shannon}}^i & = -\frac{1}{\log L_i}\sum_{k=a_i}^{b_i} \bar{P}_i[k] \log \bar{P}_i[k] 
\end{align}
where $L_i$ is the length of sequence $[a_i,b_i]$ representing the range of the frequency band.  
The normalised spectral density $\bar{P}_i[k]$ is calculated as
$\bar{P}_i[k]=P[k]/\sum_{k=a_i}^{b_i} P[k]$.
For these two entropy features, the PSD estimate $P[k]$ is either the periodogram in
\eqref{eq:2}, the Welch periodogram in \eqref{eq:8}, or the robust estimate in
\eqref{eq:9}. 
This option is set with the parameter "feat_params_st.spectral.method" as either
"'periodogram'", "'PSD'" (the default), or "'robust-PSD'".  

Spectral edge frequency is defined as the frequency $f_{\textrm{SEF}}$ that contains $d$\%
(default 95\%, parameter "feat_params_st.spectral.SEF") of the spectral energy; that is,
we find $f_{\textrm{SEF}}$ that satisfies the relation
\begin{equation}
  \label{eq:12}
  \frac{d}{100} = \frac{\sum_{k=a_1}^{f_{\textrm{SEF}}} P[k]}{\sum_{k=a_1}^{b_4} P[k]}.  
\end{equation}
As for the spectral entropy measures, the PSD estimate $P[k]$ can be either one of the 3
estimates in \eqref{eq:2}, \eqref{eq:8}, and \eqref{eq:9}, set with the parameter
"feat_params_st.spectral.method".

Spectral difference is measured as the difference between consecutive time-slices of the
spectrogram (a time-varying spectral estimate).   The spectrogram, with window $w[n]$, is
defined as
\begin{equation}
  \label{eq:13}
  S[n,k]=\left| \sum_{m=0}^{N-1} x[m]w[m-n]\ee^{-\jj 2\uppi km/N} \right|^2
\end{equation}
and the difference is calculated as 
\begin{equation}
  \label{eq:14}
  F_{\textrm{specdiff}} = \textrm{median}\left\{ \frac{1}{L_i} \sum_{k=a_i}^{b_i} \left|
        \bar{S}^i[n,k] - \bar{S}^i[n+1,k] \right|^2 \right\}
\end{equation}
where $\bar{S}^i[n,k]$ is normalised to the maximum spectrogram value
\begin{equation}
  \label{eq:15}
  \bar{S}^i[n,k]=\frac{S[n,k]}{\underset{n \in [0,N-1]; k \in [a_i,b_i]}{\max} S[n,k]}.  
\end{equation}

And lastly, we include fractal dimension as a spectral feature because of its relation
to spectral shape \cite{Higuchi1988}. 
We present two methods to estimate fractal dimension \cite{Higuchi1988,Katz1988}.  
The first method, proposed by Higuchi \cite{Higuchi1988} and set with parameter
"feat_params_st.fd.method='higuchi'", generates an estimate of curve length $C_m(q)$ at
different scale values $q$,
\begin{equation}
  C_m(q) = \frac{(N-1)}{\lfloor (N-m)/q \rfloor q^2}  
  \sum_{i=1}^{\lfloor (N-m)/q \rfloor} \big| x[m+iq] - x[m+(i-1)q] \big|  .  
\end{equation}
At each scale $q$, $C_m(q)$ is estimated over $m=1,2,\ldots,q$ and then summarised by the
mean value $C(q)=1/q \sum_{m=1}^{q}C_m(q)$.   
For a self-similar and stationary process, $C(q) \propto q^{-D}$ where $D$ is the fractal
dimension \cite{Higuchi1988}. 
By fitting a line to the points $(\log{q},\log{ C(q) })$ over $1\leq q \leq
q_{\textrm{max}}$, we estimate $-D$ as the slope of this line.  
To enforce an approximate linear sampling of $\log{q}$, scale values $q$ are set to
$q=1,2,3,4$ for $q\leq 4$ and $q=\lfloor 2^{(b+5)/4} \rfloor$ for $b=5,6,7,\ldots$ otherwise.  
The maximum value for $q$ is set in parameter "feat_params_st.fd.qmax" and defaults to
$q_{\textrm{max}}=6$.

The second method, proposed by Katz and set with parameter
"feat_params_st.fd.method='katz'", is defined as follows \cite{Katz1988}:
\begin{equation}
  \label{eq:25}
  D = \frac{\log(N-1)}{\log(d/l) + \log(N-1)} .  
\end{equation}
Length $l$ is defined as the sum of the Euclidean distance between consecutive points
$(n,x[n])$ and $(n+1,x[n+1])$ as
\begin{equation}
  \label{eq:26}
  l=\sum_{n=0}^{N-2}\sqrt{ 1 + (x[n+1]-x[n])^2 }  
\end{equation}
Extent $d$ is defined as the maximum (Euclidean) distance from starting point $(0,x[0])$
to any other point $(n,x[n])$ as
\begin{equation}
  \label{eq:27}
  d=\underset{0\leq n \leq N-1}{\max} \sqrt{ n^2 + (x[n]-x[1])^2 } 
\end{equation}
% Both $l$ and $d$ are Euclidean distance measures between points $(n,x[n])$
% \cite{Katz1988,Polychronaki2010}. 
Note that \eqref{eq:26} and \eqref{eq:27} differ to the definition in the often-cited
interpretation by Esteller \emph{et al.}
\cite{Esteller2001a} which defines the distance measure on the 1-dimensional $x[n]$ and
not on the 2-dimensional $(n,x[n])$ as originally intended \cite{Katz1988}.

\subsection{Connectivity}
\label{sec:connectivity}
 
We implement the brain symmetry index (BSI), which measures symmetry between hemispheres,
according to the specifications in \cite{VanPutten2007a}.
First, we estimate the PSD $P_m[k]$ for the $m$th channel of the EEG $x_m[n]$, for
$m=1,2,\ldots,M$. 
For $P_m[k]$, we can use either the periodogram in \eqref{eq:2}, the Welch periodogram in
\eqref{eq:8}, or the robust-PSD estimate in \eqref{eq:9}, by setting the parameter
"feat_params_st.connectivity.method" to either "'periodogram'", "'PSD'" (default), or
"'robust-PSD'".  
Left-hemisphere channels are ordered from $m=1,2,\ldots,M/2$ and right-hemisphere channels
ordered for $m=M/2+1,M/2+2,\ldots,M$. 
Next, we generate two PSDs as the mean PSD over all channels for each hemisphere; for example,
for the left hemisphere, 
\begin{equation*}
  P_{\textrm{left}}[k] = \frac{1}{M/2}\sum_{m=1}^{M/2} P_m[k] 
\end{equation*}
where $P_m[k]$ represents the PSD estimate for the $m$th channel. 
A similar process produces $P_{\textrm{right}}[k]$ for the right channels.  
The symmetry measure quantifies the differences in two PSDs for the $i$th frequency band
\begin{equation}
  \label{eq:19}
  C_{\textrm{BSI}}^i = \frac{1}{L_i}\sum_{k=a_i}^{b_i} \left|\frac{ P_{\textrm{left}}[k] -%
    P_{\textrm{right}}[k]}{ P_{\textrm{left}}[k] + P_{\textrm{right}}[k]} \right|
\end{equation}
where $[a_i,b_i]$ is the frequency range for the $i$th band and $L_i=b_i -a_i$.  

Another measure of hemisphere connectivity correlates the signal envelope $e_i[n]$ in
\eqref{eq:7} for the $i$th frequency band between channels and across the hemispheres. 
Channels are grouped into pairs based on their regional location; for example, frontal
channels are paired as $(F3,F4)$, central channels as $(C3,C4)$, and so on. 
Correlation coefficients (Pearson) are calculated for the $m$th pair $c_i(m)$, for
$m=1,2,\ldots,M/2$.  The median is used to summarise over all pairs:
\begin{equation}
  \label{eq:20}
  C_{\textrm{corr}}^i = \textrm{median}[c_i(m)].  
\end{equation}

A global coherence measure is another approach to quantifying connectivity between regions
across hemispheres. 
Coherence is calculated between channel pairs $x[n]$ and $y[n]$ as
\begin{equation}
  \label{eq:21}
  C_{xy}[k] =\frac{ |S_{xy}[k]|^2}{P_{xx}[k] P_{yy}[k] }
\end{equation}
where $P_{xx}[k]$ (and $P_{yy}[k]$) is the auto PSD of $x[n]$ (and $y[n]$) using either
the Welch periodogram in \eqref{eq:8} or the robust-PSD in \eqref{eq:9}.  
The cross-PSD for $x[n]$ and $y[n]$, is calculated as
\begin{equation}
  \label{eq:22}
  S_{xy}[k]=\frac{1}{LMUf_s}\sum_{l=0}^{L-1}\left( \sum_{n=0}^{M-1} x[n]w[n-lK]\ee^{-\jj 2\uppi kn/M}%
  \sum_{n'=0}^{M-1} y[n']w[n'-lK]\ee^{\jj 2\uppi kn'/M} \right).  
\end{equation}
for the cross-Welch periodogram.
The equivalent cross robust-PSD version replaces the mean operation over $l$, that is
$1/L \sum_{l=0}^{L-1}$, with the median operator. 
Both the auto and cross PSDs estimates are set using the parameter
"feat_params_st.connectivity.method='PSD'" (default) or
"feat_params_st.connectivity.method='robust-PSD'". 
Here, we assume that $y[n]$ and $w[n]$ are real-valued functions. 
Three features are used to summarise the coherence function $C_{xy}[k]$:
\begin{align}
  \label{eq:23}
  C_{\textrm{coh\_mean}}^i & = \frac{1}{L_i}\sum_{k=a_i}^{b_i} C_{xy}[k]    \\
  C_{\textrm{coh\_max}}^i & = \underset{k \in [a_i,b_i]}{\max} C_{xy}[k] \\
  C_{\textrm{coh\_freq\_max}}^i & = \underset{k \in [a_i,b_i]}{\arg\!\max}\; C_{xy}[k] 
\end{align}
Similar to correlation in \eqref{eq:20}, coherence features are estimated between the $m$th channel
pairs and then summarised by the median value over all channel pairs.

To eliminate spurious coupling caused by inaccuracies in the coherence measure, we have
implemented a null-hypothesis testing process to better estimate zero coherence
\cite{Prichard1994,Faes2004}. 
This approach generates a lower threshold by assessing the likelihood that a coherence
measure represents either zero coherence (the null hypothesis) or non-zero coherence. 
An empirical probability distribution is generated from multiple uncoupled signals and is
used to represent the null hypothesis that coherence is due to chance and not
significantly different to zero.  
Here, we use the Fourier-transform shuffling method to generate these
uncoupled, surrogate signals \cite{Prichard1994,Faes2004}.  
The procedure, with a slight modification, is a follows:
\begin{enumerate}
\item compute the surrogate signal $u[n]$ for $x[n]$ (of length-$N$) as follows:
  \begin{enumerate}
  \item generate length-$N$ random phase $\varphi[k]$ from a uniform distribution in the
    range $[-\uppi,\uppi]$; enforce conjugate symmetry on $\varphi[k]$ with $\varphi[0]=0$
    and, if $N$ is even, $\varphi[N/2]=0$;
  
  \item multiply the magnitude spectrum of $x[n]$ by the random phase and Fourier
    transform back to the time domain
    \begin{equation*}
      u[n] = \frac{1}{N} \sum_{k=0}^{N-1} \left| X[k] \right| \ee^{\jj \varphi[k]}
      \ee^{\jj 2\uppi kn/N} ;
    \end{equation*}
  \end{enumerate}
\item generate $v[n]$ from $y[n]$ using a similar process;
\item estimate the coherence function for $u[n]$ and $v[n]$
  \begin{equation}
    \label{eq:28}
    C_{uv}[k] =\frac{ |S_{uv}[k]|^2}{P_{xx}[k] P_{yy}[k] }
  \end{equation}
  similar to \eqref{eq:21} but using $P_{xx}[k]$ and $P_{yy}[k]$ from signals $x[n]$ and
  $y[n]$;
\item iterate this process $N_{\textrm{iter}}$ times to generate the matrix
  $\mathbf{C}_{uv}[k]$ of dimension $N_{\textrm{iter}} \times N$;
\item let $T_{\textrm{lower}}[k]$ equal the $100(1-\alpha)$th percentile of
  $\mathbf{C}_{uv}[k]$ for each value of $k$, to determine if $C_{xy}[k]$ is statistical
  significant for $p<\alpha$;
\item generate coherence $C_{xy}[k]$ between $x[n]$ and $y[n]$ and threshold: set
  $C_{xy}[k]=0$ for $C_{xy}[k]<T_{\textrm{lower}}[k]$.  
\end{enumerate}
Parameters $N_{\textrm{iter}}$ and $\alpha$ are set by
"feat_params_st.connectivity.coherence_surr_data" (default 100) and
"feat_params_st.connectivity.coherence_surr_alpha" (default 0.05). 
To calculate coherence without this zero-coherence estimation procedure, set
"feat_params_st.connectivity.coherence_surr_data=0". 
The modification of the method here, not in the original procedure
\cite{Prichard1994,Faes2004}, is to use $P_{xx}[k]$ instead of $P_{uu}[k]$ in
\eqref{eq:28}. We base this on the assumption that because the magnitude spectrum for
$u[k]$ and $x[k]$ are equal, therefore PSDs should also be equal.
The whole procedure to generate the threshold $T_{\textrm{lower}}[k]$ can be slow and this
modification reduces computational time by approximately one-half.

For all connectivity measures using the PSD, the window type and overlap for the PSDs can
be set using the parameters "feat_params_st.connectivity.PSD_window" and
"feat_params_st.connectivity.overlap".

\subsection{Burst and Inter-burst Intervals}
\label{sec:burst-inter-burst}

The EEG of preterm infants shows a discontinuous pattern with short-duration bursts of
activity alternating with longer quiescent periods.  
As the infant maturates, the burst periods become longer and quiescent periods shorter so
that eventually the EEG becomes continuous as the infant approaches term age.  
Discontinuous activity, which predominates in prematurity, consists of intermittent bursts
against a background pattern of low-amplitude activity known as inter-burst intervals
(IBI). 
To quantify this burst--inter-burst pattern, a burst detection method is needed to first
detect the bursts before summarising the bursting pattern.
We use the burst-detection algorithm proposed by O'~Toole et al. \cite{OToole2017}.  
The algorithm is freely available at \url{https://github.com/otoolej/burst_detector} and
is used to generate the following features.  

The burst detection algorithm estimates the burst annotation $b[n]$, with $b[n]=0$ for
inter-bursts and $b[n]=1$ for bursts. 
Summary measures of temporal evolution of the bursting pattern include burst percentage,
calculated as
\begin{equation}
  \label{eq:24}
  B_{\textrm{burst\%}} = \frac{100}{N} \sum_{n=0}^{N-1} b[n] 
\end{equation}
and burst number $B_{\textrm{burst\#}}$, defined as the number of detected bursts over $b[n]$.  
Similarly, summary measures of the inter-burst pattern includes $B_{\textrm{IBI\_max}}$, the maximum
duration of all IBIs, and $B_{\textrm{IBI\_median}}$ the median duration
of IBIs.

\subsection{Short-time and multi-channel analysis}
\label{sec:multi-channel-long}

Amplitude, frequency, and connectivity features are estimated on a short-time basis:
features are calculated over a short duration window and this window is shifted in time;
default window length is 64 seconds with an overlap of 50\%, set with parameters
"EPOCH_LENGTH" and "EPOCH_OVERLAP". 
The median value is used to summarise these features over all analysis windows. 
For the amplitude, frequency, and bursting features, the features are estimated separately
on each channel. Again, the median value is used to summarise over all channels.

\section{Examples}
\label{sec:results}

\subsection{Removing artefacts}
\label{sec:removing-artefacts}

\begin{figure}
  \centering
  \includegraphics[width=0.8\linewidth]{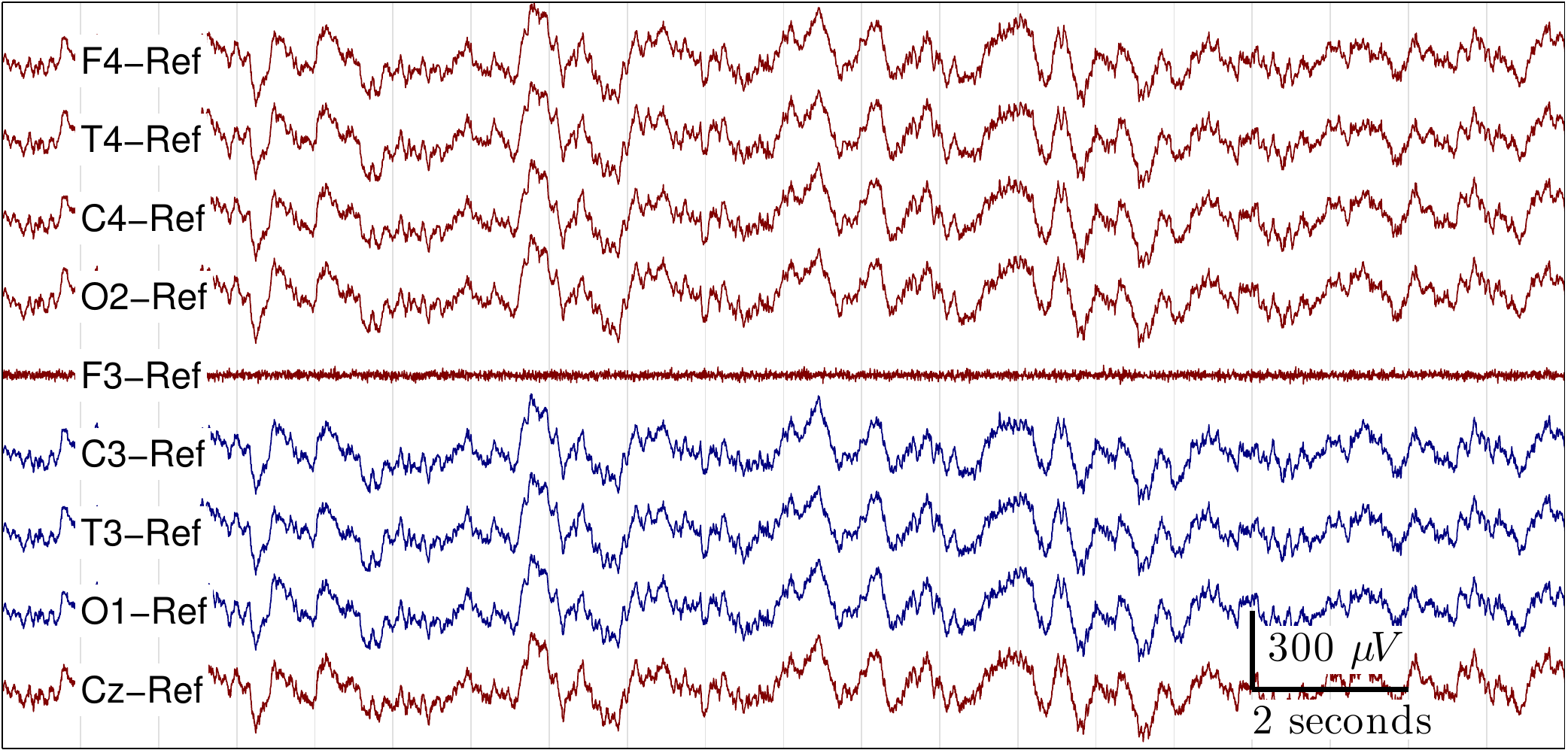}
  \includegraphics[width=0.074\linewidth]{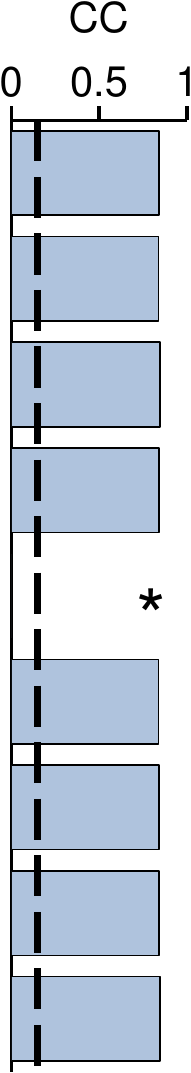}
  \caption{Referential montage of simulated EEG data (coloured Gaussian noise). All
    channels, expect for F3, are highly correlated [correlation coefficient (CC) $r>0.8$,
    right-hand side]. CCs are generated by averaging correlations between each channel and
    all other channels. 
    Channels with CCs below the given threshold ($r=0.15$, vertical dashed line) are
    removed (denoted with the $*$ symbol), as is the case here for F3. 
  }
  \label{fig:refer_mont_elect_off}
\end{figure}

\begin{figure}
  \centering
  \includegraphics[width=0.088\linewidth]{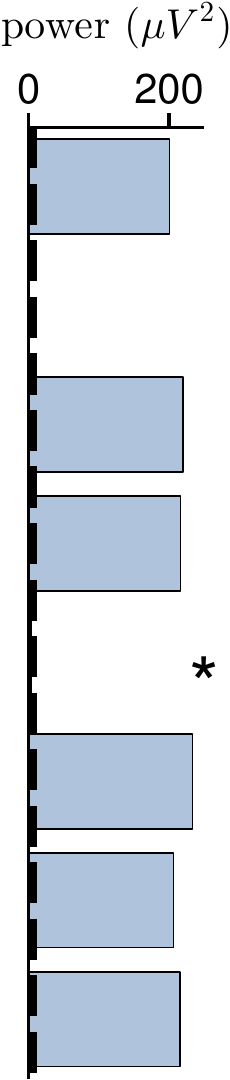}
  \includegraphics[width=0.8\linewidth]{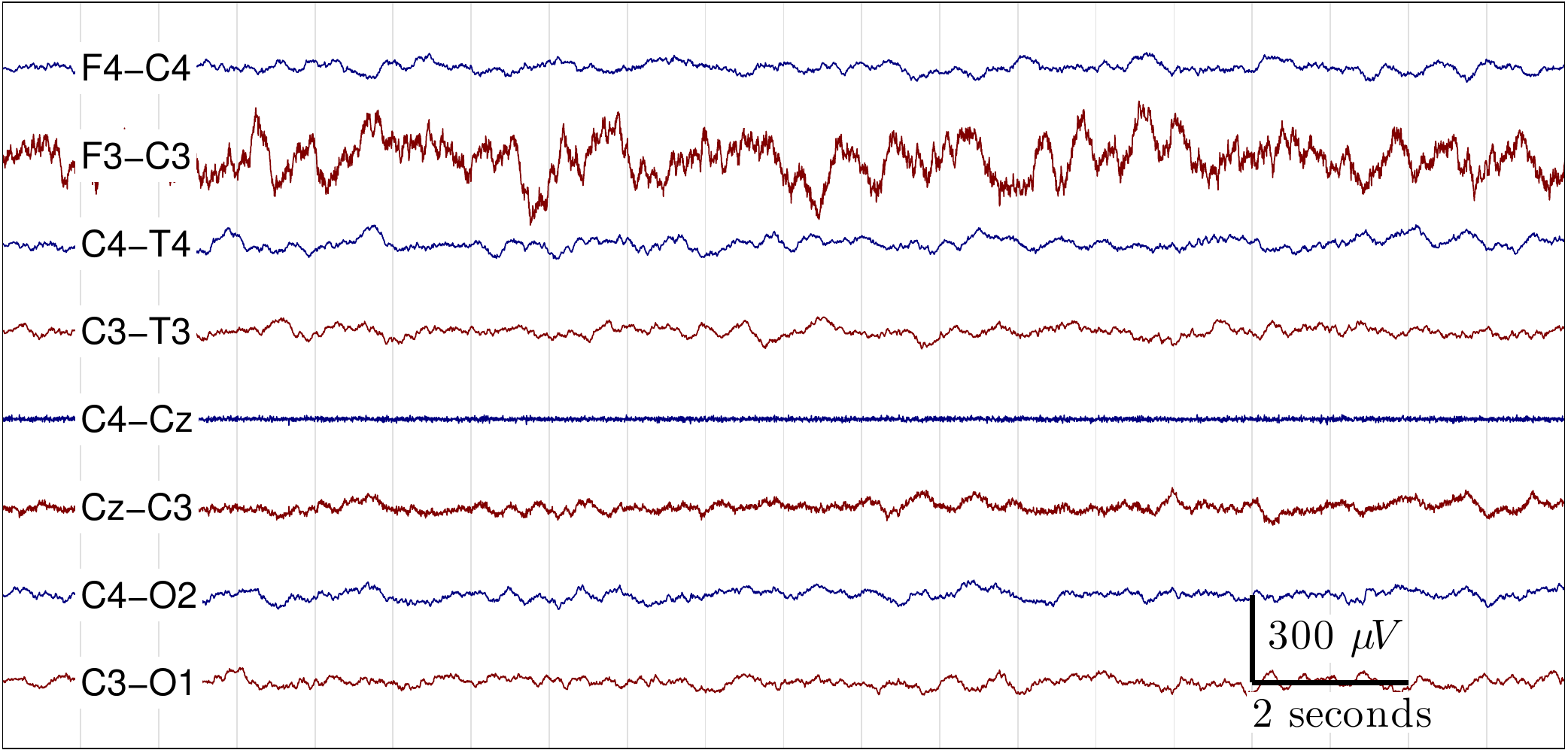}
  \caption{ Bipolar montage of EEG in Fig.~\ref{fig:refer_mont_elect_off}. Total power is
    assessed for each channel and plotted in left-hand side figure; channel F3-C3 is not
    included because F3 was removed in previous process (see
    Fig.~\ref{fig:refer_mont_elect_off}).  Channels with total power less than threshold
    (determined from median values of left and right hemisphere's separately) are denoted
    with $*$ and removed from further analysis.}
  \label{fig:bi_mont_elect_short}
\end{figure}

To describe how the artefact removal procedure works, we present an example with simulated
EEG.  The function "gen_test_EEGdata.m" generates coloured Gaussian noise as a proxy for
neonatal EEG:
\begin{lstlisting}[style=Matlab-editor]
% generates 2 minutes of EEG-like data sampled at 256 Hz:
Fs = 256;
data_st = gen_test_EEGdata(2*60,Fs,1);
\end{lstlisting}
The function returns the test data as both 9-channel referential montage
("data_st.eeg_data_ref") and the 8-channel bipolar montage ("data_st.eeg_dat").  
Next, we simulate a faulty recording on electrode F3:
\begin{lstlisting}[style=Matlab-editor]
N = size(data_st.eeg_data_ref,2);
if3 = find(strcmp(data_st.ch_labels_ref,'F3'));
data_st.eeg_data_ref(if3,:) = randn(1,N).*10;
\end{lstlisting}
Then simulate an electrode coupling between C4 and Cz
\begin{lstlisting}[style=Matlab-editor]
ic4 = find(strcmp(data_st.ch_labels_ref,'C4'));
icz = find(strcmp(data_st.ch_labels_ref,'Cz'));
data_st.eeg_data_ref(icz,:) = data_st.eeg_data_ref(ic4,:)+randn(1,N).*5;
\end{lstlisting}
and then re-generate the bipolar montage
\begin{lstlisting}[style=Matlab-editor]
[data_st.eeg_data,data_st.ch_labels] = set_bi_montage( ...
    data_st.eeg_data_ref,data_st.ch_labels_ref,data_st.ch_labels_bi);
\end{lstlisting}
The simulated EEG is displayed in Fig.~\ref{fig:refer_mont_elect_off} using the
referential montage and in Fig.~\ref{fig:bi_mont_elect_short} using the bipolar montage.  

To detect and remove these simulated artefacts, we use the "remove_artefacts.m" function:
\begin{lstlisting}[style=Matlab-editor]
eeg_art = remove_artefacts(data_st.eeg_data,data_st.ch_labels, ...  
            data_st.Fs,data_st.eeg_data_ref,data_st.ch_labels_ref);
\end{lstlisting}
which returns the data in bipolar montage with channels C4-Cz and F3-C3 replaced by "NaN"
values to indicate artefacts. 
The artefact identification process, relating to steps 1 and 2 in
Section~\ref{sec:analys-cont-multi}, is illustrated in
Figs.~\ref{fig:refer_mont_elect_off} and \ref{fig:bi_mont_elect_short}.

%% MORE examples here?

\subsection{Calculating Features}
\label{sec:calculating-features}

Features are estimated using the "generate_all_features" function, as the following
example shows.  
First, we generate 5 minutes of test data with a 64~Hz sampling rate:
\begin{lstlisting}[style=Matlab-editor]
data_st = gen_test_EEGdata(5*60,64,1);
\end{lstlisting}
Then, define the feature set as follows:
\begin{lstlisting}[style=Matlab-editor]
feature_set = {'spectral_relative_power','connectivity_BSI','rEEG_SD'};
\end{lstlisting}
Relative spectral power $S_{\textrm{relpow}}^i$ and BSI $C_{\textrm{BSI}}^i$ are defined
in equations \eqref{eq:1001} and \eqref{eq:19}, respectively; standard deviation of rEEG
$R_{\textrm{sd}}^i$ is defined in Section~\ref{sec:amplitude}. 
We can also use the parameter file "neural_parameters.m" to define the feature set. 
The full list of features are in Tables~\ref{tab:feats_spec}--~\ref{tab:feats_ibi}. 
And lastly, we estimate the features using "generate_all_features":
\begin{lstlisting}[style=Matlab-editor]
feat_st = generate_all_features(data_st,[],feature_set);
\end{lstlisting}
Features are returned as a structure, with 1 feature for each of the 4 frequency bands.  
\begin{lstlisting}[style=Matlab-editor]
feat_st = 
    spectral_relative_power: [0.8672 0.0913 0.0350 0.0243]
           connectivity_BSI: [0.0614 0.0489 0.0455 0.0505]
                    rEEG_SD: [3.1808 6.9492 4.5940 3.3538]
\end{lstlisting}

\begin{figure}
  \centering
    \labellist
    \small\hair 2pt
    \pinlabel \textbf{\labfont A.} at 1 280
    \endlabellist
    \includegraphics[width=0.8\linewidth]{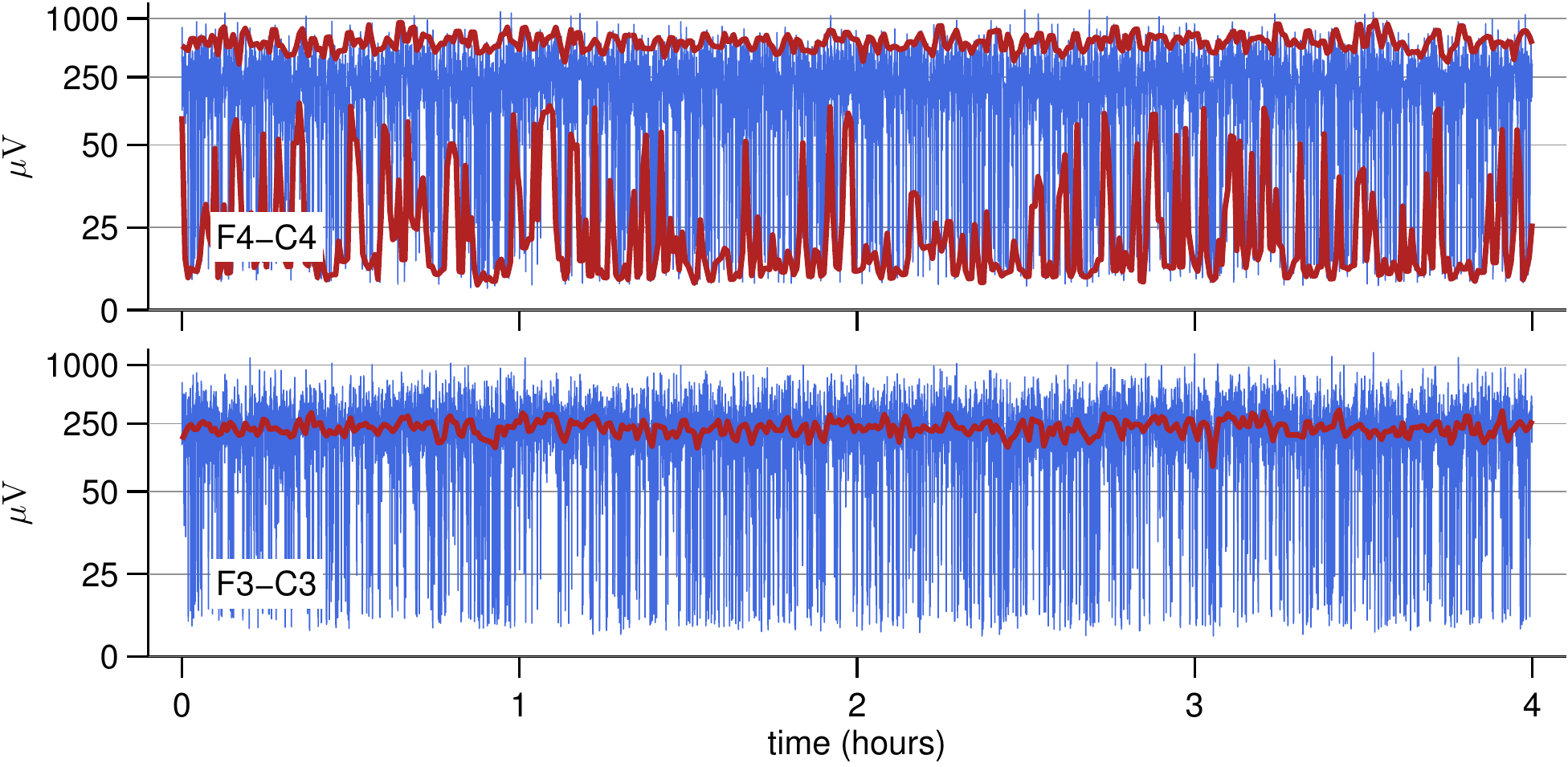}\\ \vspace{1em}\qquad
    \labellist
    \small\hair 2pt
    \pinlabel \textbf{\labfont B.} at -45 280
    \endlabellist
    \includegraphics[width=0.72\linewidth]{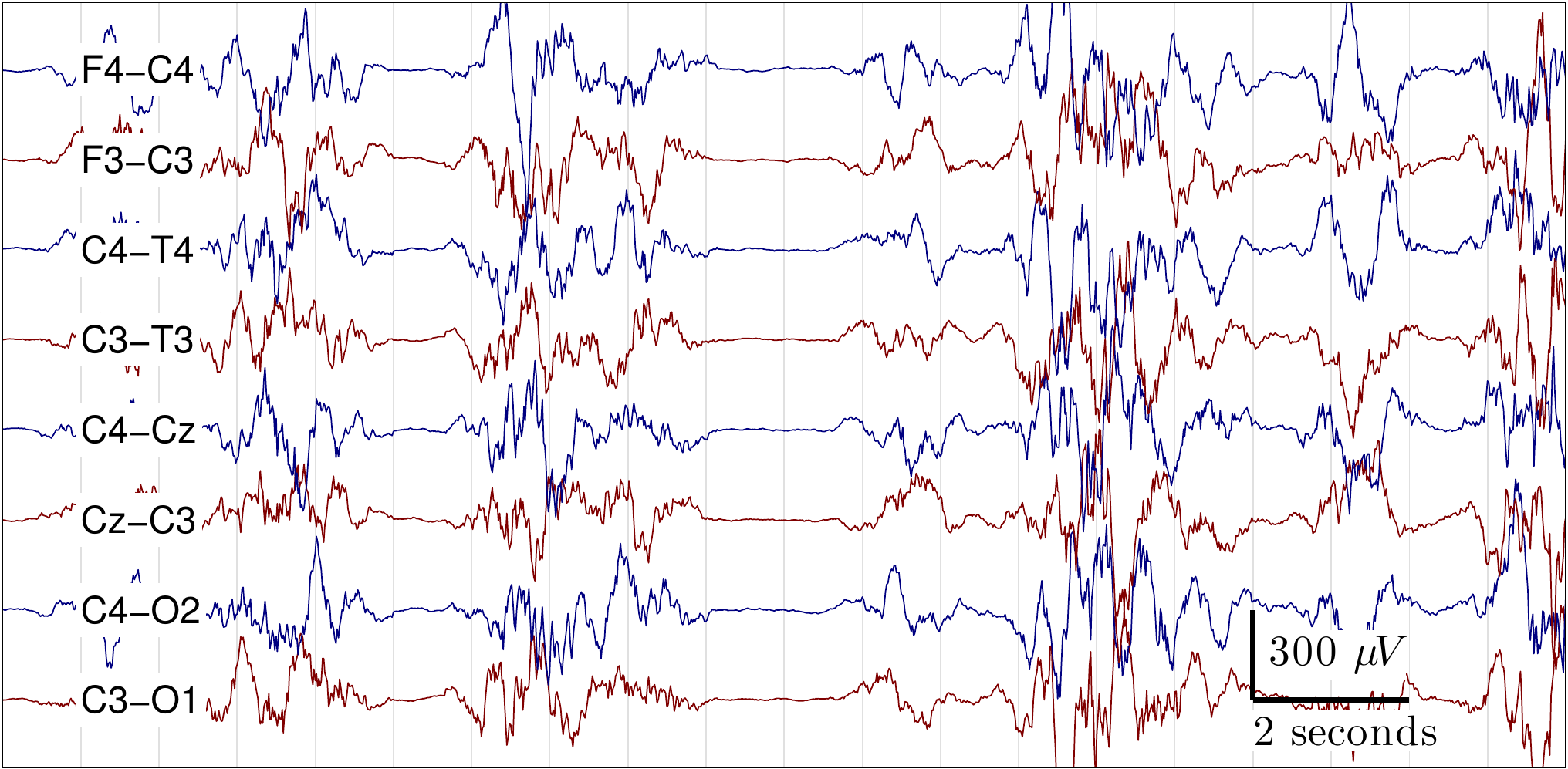}
  \caption{Estimating features with range-EEG (rEEG) using simulated EEG. A: 2-channels
    of rEEG (blue lines) over a 4-hour period. Red lines in top (A) represent upper- and
    lower-margins and median in bottom (A).  B: 20 seconds of simulated preterm.    
    EEG.}
  \label{fig:rEEG_egs}
\end{figure}

For the next example we calculate features of the rEEG.   Simulated EEG, intended to
resemble a discontinuous trace of preterm EEG, is generated as
\begin{lstlisting}[style=Matlab-editor]
data_st = gen_test_EEGdata(duration,Fs,1,1);
\end{lstlisting}
with "duration=4*60*60" (4 hours of EEG) and "Fs=64".  We then band-pass filter this
simulated EEG with a 1--20~Hz filter by setting
\begin{lstlisting}[style=Matlab-editor]
feat_params_st.rEEG.freq_bands = [1 20];
\end{lstlisting}
in the parameter file ("neural_parameters.m").  Fig.~\ref{fig:rEEG_egs}(B) shows a
20-second epoch of the 4-hour multichannel EEG.  We can view the rEEG by plotting the
second output argument from the "rEEG" function
\begin{lstlisting}[style=Matlab-editor]
[~,reeg_left] = rEEG(data_st.eeg_data(7,:),Fs);
[~,reeg_right] = rEEG(data_st.eeg_data(8,:),Fs);
\end{lstlisting}
where "reeg_left" and "reeg_right" represent the rEEG generated for channels F3-C3 and
F4-C4. 
These signals are plotted in Fig.~\ref{fig:rEEG_egs}(A). 
Next, assuming we want to estimate median, upper- and lower-margins of rEEG
($R_{\textrm{median}}$, $R_{\textrm{lower}}$, and $R_{\textrm{upper}}$ as defined in
Section~\ref{sec:amplitude}), we do as follows:
\begin{lstlisting}[style=Matlab-editor]
feats = generate_all_features(data_st,{'F3-C3','F4-C4'}, ...  
         {'rEEG_lower_margin','rEEG_upper_margin','rEEG_median'});
\end{lstlisting}
For the example in Fig.~\ref{fig:rEEG_egs}, we find median value of $R_{\textrm{median}}=228 \mu V$,
and lower--upper margins of $[15$--$578]\,\mu V$.

% Simulated rEEG in Fig.~\ref{fig:rEEG_egs}.  Lower margin: 15 $\mu$V; upper margin: 578
% $\mu$V; and median 228 $\mu$V.  (median over the 4 hours).  Using 1--20 Hz filter on the EEG.  

\section{Discussion}
\label{sec:conclusions}

The presented feature set contains commonly-used features for quantitative analysis of
newborn EEG.
The set is implement in Matlab and freely available as open source code to use and modify
as required \cite{OToole2016c}.
The features are clearly defined with all implementation details.  
Research publications often omit these details
yet they are essential for reproducible research and for comparing
results across independent studies. 
For example, power in the delta frequency band will depend on a number of factors---such
as filtering (if applied, and if so what type of filter used), power spectral density
estimation (what type of estimate and what parameters used), and how to calculate the
power (either sum over range in the frequency domain or energy of filtered signal in the
time domain).  
Different implementations will give different estimates of the delta power, therefore
undermining comparisons across studies.
The availability of a common feature set, such as the one presented here, will enable
both direct comparisons and reproducible research.  

Our goal was to present quantitative features commonly used in newborn EEG
\cite{scher1994comparisons,Duffy2003,Paul2003,Scher2004a,West2006,Victor2005b,Pereda2006,Okumura2006,Korotchikova2009,Gonzalez2011,Thorngate2013,Inder2003,Korotchikova2011,Williams2012,Saito2013,Schumacher2013a,Burdjalov2003,Niemarkt2011,O'Reilly2012,Meijer2014,Shany2014,Saji2015,Lloyd2016,OToole2016b}
and not include an exhaustive list of all possible features.  
For example, we restricted the set to stationary features and implemented a short-time
procedure to accommodate for the non-stationary aspects of EEG.
Although non-stationary methods have been used to analyse newborn EEG
\cite{Stevenson2011,Mesbah2012,Omidvarnia2013b,Boashash2013}, we do not include these here
because of the following: first, these nonstationary methods are often developed for
specific applications, such as detecting seizures \cite{Stevenson2011}; second, their
complexity limits their use, as typically a detailed-level of understanding of these
methods is required before implementation \cite{Boashash2013}; and third, in many cases
the assumption of short-time stationarity may be acceptable for the required
application, for example in seizure detection \cite{Temko2010}. 
Future iterations could expand the scope of feature set, to include features such as more
advanced connectivity measures \cite{Omidvarnia2013} or methods to quantify activity
cycling in preterm newborns \cite{0967-3334-35-7-1493}.  

In conclusion, the NEURAL software package provides a common platform for quantitative
analysis of multi-channel EEG for newborn infants.
The software will assist reproducible research and consistency across independent studies.
Many of the features, with the exception of the burst and inter-burst features, may also
be of use for paediatric and adult EEG.  
These quantitative features can also be used to develop automated methods for newborn,
such as automated detection of seizures \cite{Stevenson2011,Temko2010} or estimation of
brain maturation \cite{OToole2016b}.

%-------------------------------------------------------------------------------
% ACKNOWLEDGEMENTS
%-------------------------------------------------------------------------------
\section{Acknowledgements}
This work was supported by Science Foundation Ireland (research-centre award
INFANT-12/RC/2272 and investigator award 15/SIRG/3580). 
We thank Brian Murphy for help with testing the Matlab code.

%-------------------------------------------------------------------------------
% APPENDIX
%-------------------------------------------------------------------------------
\appendix
\section{Moments for Stationary Random Variable}
\label{sec:moments-stat-rand}

The first 4 moments (mean, standard deviation, skewness, and kurtosis) for random variable
$x[n]$ are estimated as follows:
mean %($\textrm{mean}(x[n])$)
\begin{equation*}
  \bar{x} = \frac{1}{N}\sum_{n=0}^{N-1} x[n],
\end{equation*}
standard deviation %($\textrm{std}(x[n])$)
\begin{equation*}
  s = \sqrt{\frac{1}{N-1}\sum_{n=0}^{N-1} \left| x[n] - \bar{x} \right|^2 },
\end{equation*}
skewness %($\textrm{skew}(x[n])$)
\begin{equation*}
    m_3 = \frac{{\frac{1}{N}\sum_{n=0}^{N-1} \left| x[n] - \bar{x} \right|^3 }}{s^3},
\end{equation*}
and kurtosis %($\textrm{kurt}(x[n])$)
\begin{equation*}
  m_4 = \frac{{\frac{1}{N}\sum_{n=0}^{N-1} \left| x[n] - \bar{x} \right|^4 }}{s^4} .  
\end{equation*}

\section{List of Features}
\label{sec:list-features}

\begin{table}[htpb]
  \centering
  \caption{Amplitude features with Matlab code name. 
    Features are calculated for each frequency band.}
  \begin{tabulary}{0.9\linewidth}{lL}
    \toprule
    name & description \\
    \midrule
{\lstinline[style=Matlab-editor]!amplitude\_total\_power!}  %
         & time-domain signal: total power \\
{\lstinline[style=Matlab-editor]!amplitude\_SD!}  %
            & time-domain signal: standard deviation                                             \\
{\lstinline[style=Matlab-editor]!amplitude\_skew!}  %
            & time-domain signal: skewness                                                  \\
{\lstinline[style=Matlab-editor]!amplitude\_kurtosis!}  %
        & time-domain signal: kurtosis                                                  \\
{\lstinline[style=Matlab-editor]!amplitude\_env\_mean!}  %
       & envelope: mean value                                                          \\
{\lstinline[style=Matlab-editor]!amplitude\_env\_SD!}  %
         & envelope: standard deviation                                              \\
{\lstinline[style=Matlab-editor]!rEEG\_mean!}  %
                 & range EEG: mean                                                               \\
{\lstinline[style=Matlab-editor]!rEEG\_median!}  %
               & range EEG: median                                                             \\
{\lstinline[style=Matlab-editor]!rEEG\_lower\_margin!}  %
        & range EEG: lower margin (5th percentile)                                      \\
{\lstinline[style=Matlab-editor]!rEEG\_upper\_margin!}  %
        & range EEG: upper margin (95th percentile)                                     \\
{\lstinline[style=Matlab-editor]!rEEG\_width!}  %
                & range EEG: upper margin - lower margin                                        \\
{\lstinline[style=Matlab-editor]!rEEG\_SD!}  %
                   & range EEG: standard deviation                                                 \\
{\lstinline[style=Matlab-editor]!rEEG\_CV!}  %
                   & range EEG: coefficient of variation                                           \\
{\lstinline[style=Matlab-editor]!rEEG\_asymmetry!}  %
            & range EEG: measure of skew about median                                       \\
 \bottomrule
  \end{tabulary}
  \label{tab:feats_reeg}
\end{table}

\begin{table}[htpb]
  \centering
  \caption{Spectral features with Matlab code name.}
  \begin{tabulary}{0.9\linewidth}{lL}
    \toprule
    name & description \\
    \midrule
{\lstinline[style=Matlab-editor]!spectral\_power!}$ ^\dagger$  %
         & spectral power: absolute   \\
{\lstinline[style=Matlab-editor]!spectral\_relative\_power!}$ ^\dagger$  %
         & spectral power: relative (normalised to total spectral power) \\
{\lstinline[style=Matlab-editor]!spectral\_flatness!}$ ^{\dagger}$ %
         & spectral entropy: Wiener (measure of spectral flatness)       \\
{\lstinline[style=Matlab-editor]!spectral\_entropy!}$ ^{\dagger}$  %
         & spectral entropy: Shannon                                     \\
{\lstinline[style=Matlab-editor]!spectral\_diff!}$ ^{\dagger}$  %
        & difference between consecutive short-time spectral estimates \\
{\lstinline[style=Matlab-editor]!spectral\_edge\_frequency!}  
        & spectral edge frequency: 95\% of spectral power contained between 0.5 and $f_c$
          Hz (cut-off frequency) \\
{\lstinline[style=Matlab-editor]!FD!}$ ^{\dagger}$ & fractal dimension \\
    \bottomrule
  \end{tabulary}
  \begin{flushleft}
    $^\dagger$ feature is calculated for each frequency band.  
  \end{flushleft}
  \label{tab:feats_spec}
\end{table}

\begin{table}[htpb]
  \centering
  \caption{Connectivity features with Matlab code name. Features are calculated for each
    frequency band. }
  \begin{tabulary}{0.9\linewidth}{lL}
    \toprule
    name & description \\
    \midrule
{\lstinline[style=Matlab-editor]!connectivity\_BSI!}  %
          & brain symmetry index \\
{\lstinline[style=Matlab-editor]!connectivity\_corr!}  %
         & correlation (Pearson) between envelopes of hemisphere-paired channels \\
{\lstinline[style=Matlab-editor]!connectivity\_coh\_mean!}  %
    & coherence: mean value \\
{\lstinline[style=Matlab-editor]!connectivity\_coh\_max!}  %
     & coherence: maximum value \\
{\lstinline[style=Matlab-editor]!connectivity\_coh\_freqmax!}  %
 & coherence: frequency of maximum value \\
 \bottomrule
  \end{tabulary}
  \label{tab:feats_conn}
\end{table}

\begin{table}[htpb]
  \centering
  \caption{Inter-burst interval features with Matlab code name.}
  \begin{tabulary}{0.9\linewidth}{lL}
    \toprule
    name & description \\
    \midrule
{\lstinline[style=Matlab-editor]!IBI\_length\_max!}  %
           & burst annotation: maximum (95th percentile) inter-burst interval \\
{\lstinline[style=Matlab-editor]!IBI\_length\_median!}  %
        & burst annotation: median inter-burst interval \\                         
{\lstinline[style=Matlab-editor]!IBI\_burst\_prc!}  %
            & burst annotation: burst percentage          \\
{\lstinline[style=Matlab-editor]!IBI\_burst\_number!}  %
         & burst annotation: number of bursts \\               
 \bottomrule
  \end{tabulary}
  \label{tab:feats_ibi}
\end{table}

% ---------------------------------------------------------------------
% BIBLIOGRAPHY
% ---------------------------------------------------------------------

\end{document}